\documentclass[aps,prl,twocolumn,showkeys,superscriptaddress,nofootinbib]{revtex4}
\usepackage{graphicx}
\usepackage{capt-of}
\usepackage{amsmath}
\begin{document}
\title{Implications of a matter-radius measurement for the structure of Carbon-22}
\author{B.~Acharya}
\email{acharyab@phy.ohiou.edu}
\affiliation{Department of Physics and Astronomy, 
Ohio University, Athens, OH 45701, USA}

\author{C.~Ji}
\email{jichen@triumf.ca}
\affiliation{Department of Physics and Astronomy, 
Ohio University, Athens, OH 45701, USA}
\affiliation{TRIUMF, 4004 Wesbrook Mall, Vancouver, BC V6T 2A3, Canada\footnote{Permanent Address}}

\author{D.~R.~Phillips}
\email{phillips@phy.ohiou.edu}
\affiliation{Department of Physics and Astronomy, 
Ohio University, Athens, OH 45701, USA}

\date{\today}

\begin{abstract}
We study Borromean 2$n$-halo nuclei using effective field theory. We compute the universal scaling function that relates the mean-square matter radius of the 2$n$ halo to dimensionless ratios of two- and three-body energies. We use the experimental value of the rms matter radius of $^{22}\text{C}$ measured by Tanaka et al.~(2010)~\cite{Tanaka:2010zza} to put constraints on its 2$n$ separation energy and the $^{20}\text{C}-n$ virtual energy. We also explore the consequences of these constraints for the existence of excited Efimov states in this nucleus. We find that, for $^{22}\text{C}$ to have an rms matter radius within 1$-\sigma$ of the experimental value, the two-neutron separation energy of $^{22}\text{C}$ needs to be below 100~keV. Consequently, this three-body halo system can have an excited Efimov state only if the $^{20}\text{C}-n$ system has a resonance within 1 keV of the scattering threshold.
\end{abstract}

\keywords{Efimov states, effective field theory, few-body systems, halo nuclei}

\maketitle

In the last twenty years several nuclei where the neutron distribution extends far beyond that of the protons have been discovered. These nuclei, such as ${}^{11}$Li and ${}^{12}$Be thus have neutron ``halos" (see Refs.~\cite{Tanihata:1995yv,673808} for early reviews).
Understanding the way in which nuclear structure changes in these neutron-rich systems may provide important clues to the behavior of nuclei far from the $N=Z$ line. 

The most neutron-rich isotope of Carbon yet produced, ${}^{22}\text{C}$, has recently been identified as another example of a halo system.
Tanaka et al. measured the reaction cross-section of $^{22}\text{C}$ on a hydrogen target and, using Glauber calculations, deduced a $^{22}\text{C}$ rms matter radius of $5.4\pm 0.9$~fm~\cite{Tanaka:2010zza}. Their measurement implies that the two valence neutrons in $^{22}\text{C}$ preferentially occupy the $1s_{1/2}$ orbital and are weakly bound \cite{Tanaka:2010zza}.  This conclusion is supported by data on high-energy two-neutron removal from ${}^{22}$C~\cite{Kobayashi:2011mm}. 
Since $^{21}\text{C}$ is unbound \cite{Langevin:1985madeup}, this suggests that $^{22}\text{C}$ is an s-wave Borromean halo nucleus with two neutrons orbiting a $^{20}\text{C}$ core.

In this work, we implement such a description of ${}^{22}$C in an  effective field theory (EFT) for systems with short-range interactions. 
This EFT was developed to study few-nucleon systems with scattering length much larger than the interaction range (see Refs.~\cite{Bedaque:2002mn,Hammer:2010kp} for reviews). It was then applied to the $\alpha-n$ system in Refs.~\cite{Bertulani:2002sz,Bedaque:2003wa}, extended to study three-body systems in Ref.~\cite{Bedaque:1998kg} , and applied to 2{\emph{n}}-halo nuclei in Refs.~\cite{Canham:2008jd,Canham:2009xg}. 
At leading order (LO) in the theory the only inputs to the equations that describe a 2$n$ halo are the energies of the neutron-core resonance/bound-state, $E_{nc}$, and the $nn$ virtual bound state, $E_{nn}$, as well as the binding energy, $B$, of the halo nucleus~\footnote{The binding energy of ${}^{22}$C treated as a three-body system is equal to the two-neutron separation energy of the nucleus, $S_{2n}$.}.
All other properties of the nucleus are predicted once these (together with the core-neutron mass ratio $A$) are specified. Here we use the EFT to compute the function that describes the mean-square matter radius of an arbitrary Borromean 2$n$ halo.

We then focus on exploring properties of $^{22}$C, treated as a $2n$-halo nucleus with a $^{20}$C core. For $^{22}\text{C}$, neither $E_{nc}$ nor $B$ is well-known \cite{Audi:2002rp,Fortune:2012zzb,Coraggio:2010jz,Horiuchi:2006ds,AbuIbrahim:2007tp}. We therefore use EFT to find constraints in the $(B,E_{nc})$ plane using Tanaka et al.'s value of the rms matter radius.  A similar strategy was recently pursued in Ref.~\cite{Fortune:2012zzb}, although there a simpler model of ${}^{22}$C's structure was employed. 
The connection between the binding energy and several low-energy properties of ${}^{22}$C, including the rms matter radius, has also been explored in a three-body model by Ershov et al.~\cite{Ershov:2012fy}. 

Yamashita and collaborators investigated such correlations in halo nuclei already in 2004~\cite{Yamashita:2004pv} (see also Ref.~\cite{Frederico:2012xh} for a review). In 2011 Yamashita et al. \cite{Yamashita:2011cb} attempted to apply EFT to  analyze the experiment of Ref.~\cite{Tanaka:2010zza}.  However, as we shall discuss further below, an additional assumption was made in Ref.~\cite{Yamashita:2011cb} which renders the results of Yamashita et al. model-dependent. The results we obtain here therefore differ---both in principle and in practice---from those of Ref.~\cite{Yamashita:2011cb}.

The EFT description of three-body systems developed in Ref.~\cite{Bedaque:1998kg} also provides insights into the Efimov effect~\cite{Efimov:1970zz} (see Ref.~\cite{Braaten:2002} for a review). This phenomenon occurs in the three-body system when the two-body interaction generates a scattering length, $a$, that is large compared to its range. For sufficiently large $a$ there is a sequence of three-body states whose properties are related by discrete scale transformations. Refs.~\cite{Mazumdar:2000dg,Canham:2008jd,Canham:2009xg,Frederico:2012xh} explore the possibility of finding excited Efimov states in several 2$n$-halo nuclei, including $^{20}\text{C}$ and ${}^{22}$C. We use our constraint on the binding energy of ${}^{22}$C to discuss the possibility that such states occur there. 

\section{Halo EFT at LO} 

Effective field theory is a powerful tool to study few-body systems at low energies, because long-distance properties of the system are insensitive to  the details of the underlying short-range interactions. In the case of the effective field theory for halo nuclei (``Halo EFT") the breakdown scale of the EFT, $\Lambda_0$, is of the order of the inverse of the range of the two-body interactions, and this is much larger than $Q$, the generic low-momentum scale of the system. $Q$ can represent either the momentum of the process $p$, or $1/a$. At LO, we ignore terms suppressed by $Q/\Lambda_0$, which amounts to approximating the two-body potentials to be zero-range. The results of LO EFT calculations are, therefore, universal in the sense that they are independent of the short-distance physics. The ``range effects" can be systematically taken into account order-by-order in the $Q/\Lambda_0$ expansion~\cite{Kaplan:1998tg,Kaplan:1998we,Gegelia:1998gn,Birse:1998dk,vanKolck:1998bw,Hammer:2001gh,Ji:2010su,Ji:2011qg,
Ji:2012nj}. However, even in an LO calculation, one can estimate the uncertainty due to these neglected higher-order terms. This feature of EFT allows us to account for its theoretical uncertainty, and makes it uniquely suited to constraining $B$ and  $E_{nc}$ of $^{22}{\text C}$ from the matter-radius measurement.

\section{Faddeev Equations}
\label{sec:fadeev}
The two-body virtual energies, $E_{nx}$, are related to the scattering lengths, $a_{nx}$, by 
\begin{equation}
E_{nx} = \frac{1}{2\mu_{nx} a_{nx}^2}+...,
\end{equation} where, $x=n~\text{or}~c$,  $\mu_{nx}$ is the reduced mass of the corresponding two-body system, and the ellipses indicate higher-order corrections. Throughout this Letter, we work in units with $\hbar=c=1$. 

Following Refs.~\cite{Bedaque:1998kg,Canham:2008jd}, we write the Faddeev equations for the spectator functions, $F_x(q)$, which describe the relative motion of the spectator particle, $x$, and the center of mass of the other two particles for an s-wave three-body bound state of two neutrons of mass $m$ each, and a core of mass $Am$.
\begin{eqnarray}
F_n(q)=\dfrac{1}{2}\int_0^\infty \text{d}q'~q'^2  \int_{-1}^1 \text{d}\left(\hat{q}.\hat{q}^{\, \prime}\right)\qquad\qquad\qquad\nonumber\\ 
\lbrace[G_0^n\left(\pi(\vec{q}^{\, \prime},\vec{q}),q';B\right)+H(\Lambda)]~t_n(q';B)F_n(q')\nonumber\\
+G_0^c\left(\pi_2(\vec{q},\vec{q}^{\,\prime}),q';B\right)t_c(q';B)F_c(q')\rbrace,
\label{eq:fadeevfn}
\end{eqnarray}
\begin{eqnarray}
F_c(q)=\int_0^\infty \text{d}q'~q'^2  \int_{-1}^1 \text{d}\left(\hat{q}.\hat{q}^{\, \prime}\right)\qquad\qquad\qquad\quad\nonumber\\ 
G_0^n\left(\pi_1(\vec{q}^{\, \prime},\vec{q}),q';B\right)t_n(q';B)F_n(q'),
\label{eq:fadeevfc}
\end{eqnarray}
where the momenta $\vec{\pi}$, $\vec{\pi}_1$, and $\vec{\pi}_2$ are given by
\begin{equation}
\vec{\pi}(\vec{q},\vec{q}^{\, \prime})=\left(\frac{1}{A+1}\right)\vec{q}+\vec{q}^{\, \prime},
\end{equation}
\begin{equation}
\vec{\pi}_1(\vec{q},\vec{q}^{\, \prime})=\left(\frac{A}{A+1}\right)\vec{q}+\vec{q}^{\, \prime},
\end{equation}
and
\begin{equation}
\vec{\pi}_2(\vec{q},\vec{q}^{\, \prime})=\vec{q}+\frac{1}{2}\vec{q}^{\, \prime}.
\end{equation} The three-body Green's functions $G_0^n$ and $G_0^c$ are given by
\begin{equation}
G_0^n(p,q;B)=\left(B+\frac{A+1}{2Am}p^2+\frac{A+2}{2(A+1)m}q^2\right)^{-1},
\end{equation}
and
\begin{equation}
G_0^c(p,q;B)=\left(B+\frac{p^2}{m}+\frac{A+2}{4Am}q^2\right)^{-1}.
\end{equation} The two-body $t$-matrices are 
\begin{equation}
t_n(q;B)=\frac{A+1}{\pi Am}
\frac{1}{-\frac{1}{a_{nc}}+\sqrt{\frac{A}{A+1}\left(2mB+\frac{A+2}{A+1}q^2\right)}},
\end{equation}
and
\begin{equation}
t_c(q;B)=\frac{2}{\pi m}\frac{1}{-\frac{1}{a_{nn}}+\sqrt{mB+\frac{A+2}{4A}q^2}}.
\end{equation}

To solve the Faddeev equations, an ultraviolet cutoff, $\Lambda$, needs to be introduced to the integrals. The three-body contact interaction, $H(\Lambda)$, is then required to cancel the cutoff-dependence \cite{Bedaque:1998kg}. $H(\Lambda)$ scales as $mh(\Lambda)/\Lambda^2$ where $h(\Lambda)\sim\mathcal{O}(1)$. We have verified that one three-body counter term, added to $G_0^n$ in Eq.~\eqref{eq:fadeevfn}, is sufficient to renormalize Eqs.~\eqref{eq:fadeevfn}~and~\eqref{eq:fadeevfc}. Furthermore, all the results in this Letter can also be replicated by adding a three-body counter term to each of the Green's functions in Eqs.~\eqref{eq:fadeevfn} and \eqref{eq:fadeevfc}.

In Ref.~\cite{Canham:2008jd}, Gaussian regulators were introduced in the two-body potential to regularize the effects of the short-range interactions on the two and three-body observables at a cutoff scale $\Lambda$. This is equivalent to our zero-range approximation up to corrections of order $Q/\Lambda$ in the EFT. In Refs.~\cite{Yamashita:2004pv,Yamashita:2011cb,Frederico:2012xh} a subtraction was performed on the equations for the spectator functions. The subtraction functions were then set by assuming that the Born approximation holds for an appropriate subtraction point. This assumption is, however, not valid for the integral equations which are employed here~\cite{Afnan:2003bs,Yang:2007hb}.
 
The three-body wave function in the Jacobi representation with the neutron as the spectator is
\begin{eqnarray}
\Psi_n(p,q)=G_0^n(p,q;B)~\lbrace t_n(q;B)F_n(q)\qquad\qquad\quad\nonumber\\
+\frac{1}{2} \int_{-1}^1 \text{d}\left(\hat{p}.\hat{q}\right)t_n(\pi'_{nn}(\vec{p},\vec{q});B)F_n(\pi'_{nn}(\vec{p},\vec{q}))\nonumber\\
+t_c(\pi'_{nc}(\vec{p},\vec{q});B)F_c(\pi'_{nc}(\vec{p},\vec{q}))\rbrace,
\end{eqnarray} 
and with the core as the spectator is
\begin{eqnarray}
\Psi_c(p,q)=G_0^c(p,q;B)~\lbrace t_c(q;B)F_c(q)\qquad\qquad\qquad\nonumber\\
+\int_{-1}^1 \text{d}\left(\hat{p}.\hat{q}\right)t_n(\pi'_{cn}(\vec{p},\vec{q});B)F_n(\pi'_{cn}(\vec{p},\vec{q}))\rbrace,
\end{eqnarray} where the momenta $\vec{\pi}'_{nn}$, $\vec{\pi}'_{nc}$, and $\vec{\pi}'_{cn}$ are given by
\begin{equation}
\vec{\pi}'_{nn}(\vec{p},\vec{q})=\vec{p}-\left(\frac{1}{A+1}\right)\vec{q},
\end{equation}
\begin{equation}
\vec{\pi}'_{nc}(\vec{p},\vec{q})=\vec{p}-\left(\frac{A}{A+1}\right)\vec{q},
\end{equation}
and
\begin{equation}
\vec{\pi}'_{cn}(\vec{p},\vec{q})=\vec{p}-\frac{1}{2}\vec{q}.
\end{equation} 
The one-body form factors are then given by
\begin{eqnarray}
{\cal F}_x(k^2)=\int_0^{\infty} \mathrm{d}p~p^2\int_0^{\infty}\mathrm{d}q~q^2\int_{-1}^1 \text{d}\left(\hat{q}\cdot\hat{k}\right)\qquad\nonumber\\
\Psi_x(p,q)~\Psi_x(p,\lvert\vec{q}-\vec{k}\rvert),
\end{eqnarray} where the wave functions $\Psi_{x}$ are normalized so that ${\cal F}_{x}(0)=1$.
The mean-square distance of the neutron from the center of mass of the core and the other neutron, $\langle r_{n-nc}^2 \rangle$, can be extracted from
\begin{equation}
{\cal F}_n(k^2)=1-\frac{1}{6}k^2 \langle r_{n-nc}^2 \rangle+ \ldots,
\end{equation} 
and the mean-square distance of the core from the center of mass of the neutrons, $\langle r_{c-nn}^2 \rangle$, from
\begin{equation}
{\cal F}_c(k^2)=1-\frac{1}{6}k^2 \langle r_{c-nn}^2 \rangle+ \ldots.
\end{equation}

\section{A two-neutron halo nucleus with a point-like core} 
\label{sec:pointcore}
Based on geometrical arguments,  we obtain the following formula for the mean-square matter radius of a two-neutron halo in the point-like core approximation, $\langle r_0^2 \rangle$: 
\begin{equation}
\langle r_0^2 \rangle = \frac{2(A+1)^2}{(A+2)^3} \langle r_{n-nc}^2 \rangle +\frac{4A}{(A+2)^3}  \langle r_{c-nn}^2 \rangle.
\label{eq:pointlike}
\end{equation}
At LO in Halo EFT, the quantity $mB\langle r_0^2\rangle$ depends on all the variables featuring in the Faddeev equations: $E_{nn}$, $E_{nc}$, $B$ and $A$. But, being dimensionless itself, it can only depend on dimensionless ratios of these four parameters. Thus it is convenient to define the function $f \left(E_{nn}/B,E_{nc}/B;A\right)$, as~\cite{Yamashita:2004pv}:
\begin{equation}
mB\langle r_0^2\rangle \equiv f \left(\frac{E_{nn}}{B},\frac{E_{nc}}{B};A\right).
\label{eq:definef}
\end{equation}

\begin{figure}[h]
\begin{center}
\includegraphics[width=0.48\textwidth]{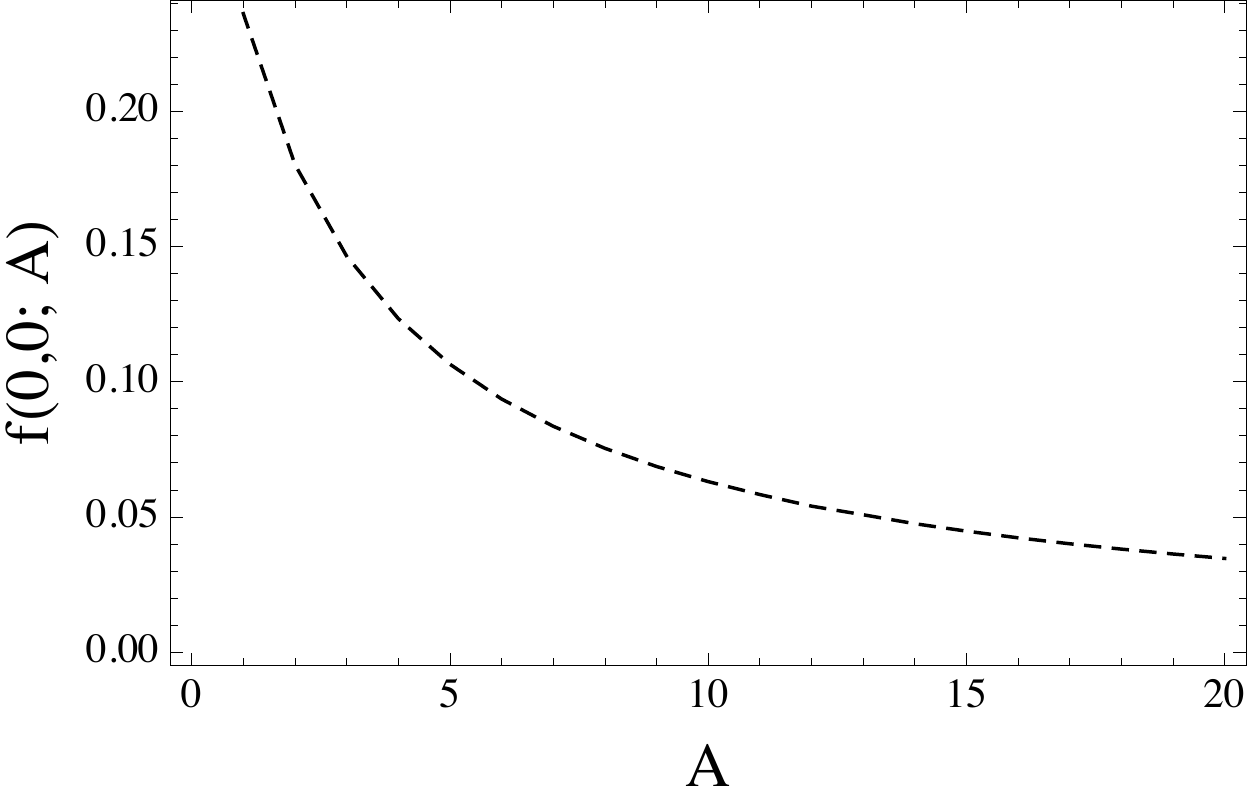}
\caption{The dimensionless function $f(0,0;A)$, defined by Eq.~\eqref{eq:definef}, versus $A$.}
\label{fig:f00avsa}
\end{center}
\end{figure}

We calculated $mB\langle r_0^2 \rangle$ at the unitary limit $E_{nn}=E_{nc}=0$. It is plotted in Figure~\ref{fig:f00avsa} as a function of $A$. These results are in qualitative agreement with those of Ref.~\cite{Yamashita:2004pv}, but our value of $f$ is lower, which means that Yamashita et al. overpredict the unitary limit $\langle r_0^2 \rangle$ for a given $B$. The discrepancy is approximately 15\% for $A=20$. 

\section{Implications for  $^{22}\text{C}$}
\label{sec:c22}

\begin{figure}[h]
\begin{center}
\includegraphics[width=0.48\textwidth]{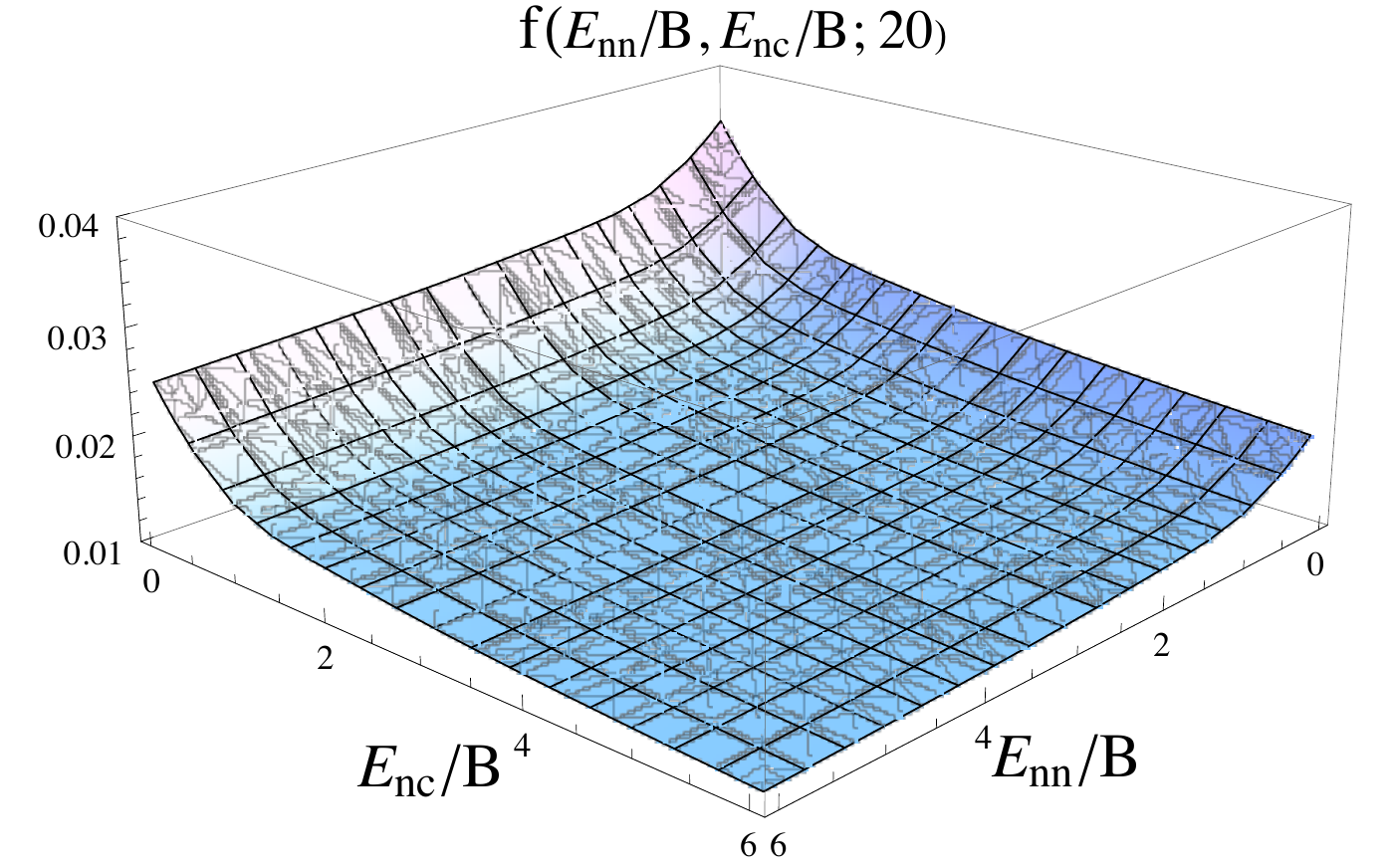}
\caption{$f \left(E_{nn}/B,E_{nc}/B;20\right)$ versus $\left(E_{nn}/B,E_{nc}/B\right)$.}
\label{fig:f3d}
\end{center}
 \end{figure}

The function $f$ can be calculated for any value of $A$, but specializing now to the case of interest for ${}^{22}$C, 
Fig.~\ref{fig:f3d} shows a three-dimensional plot of $f \left(E_{nn}/B,E_{nc}/B;20\right)$ in the $(E_{nn}/B,E_{nc}/B)$ plane. The disagreement with the results of Ref.~\cite{Yamashita:2011cb} appears to be worse at finite values of $E_{nc}$ and $E_{nn}$ than the 15\% we found in the limit $E_{nn}=E_{nc}=0$.

Fig.~\ref{fig:f3d} gives us the results we need in order to set a model-independent constraint on the binding energies of ${}^{21}$C and ${}^{22}$C. First, though, we must account for the fact that, when applied to this system, Halo EFT only reliably predicts the difference between the ${}^{22}$C matter radius and that of ${}^{20}$C. 
We account for the finite spatial extent of the core by including that effect in our expression for the 
mean-square matter radius of the two-neutron halo:
\begin{equation}
\langle r^2 \rangle = \langle r_0^2 \rangle+\frac{A}{A+2}\langle r^2 \rangle_{\rm core}.
\label{eq:msradius}
\end{equation}
Here $\langle r^2 \rangle_{\rm core}$ is the mean-square radius of the core, which we take from the $^{20}\text{C}$ rms radius of $(2.98\pm0.05)$~fm measured by Ozawa et al. \cite{Ozawa:2000gx}. In subsequent calculations we also use the value of $E_{nn}$ obtained from $a_{nn}=\left(-18.7\pm0.6\right)$~fm \cite{Trotter:1999madeup}. 

To calculate the cutoff of our EFT, we approximate the range of the neutron-core interaction by the size of the  $^{20}\text{C}$ rms radius. We then estimate the relative error of our calculation by $\sqrt{mE_{nn}}/\Lambda_0$,  $\sqrt{2mE_{nc}}/\Lambda_0$ or $\sqrt{2mB}/\Lambda_0$, whichever is the largest. 
The spectrum of ${}^{20}$C has also been measured~\cite{Stanoiu:2008zz}. It contains one bound $2^+$ state which lies 1.588 MeV above the ground state. We have not used this energy scale in assessing the breakdown of Halo EFT for ${}^{22}$C, since the $2^+$ state can affect s-wave scattering processes only via higher-dimensional operators which do not enter the calculation at next-to-leading order.  

More generally, one might be concerned about the impact of neutrons in d-wave states on the structure of ${}^{22}$C. The LO EFT being used here does not preclude the existence of such states in either ${}^{20}$C or ${}^{22}$C, it only assumes that their primary effect on long-distance dynamics can be subsumed into the neutron-core and neutron-neutron-core contact interactions which appear in the leading-order calculation. 

\subsection{Constraints on $E_{nc}$ and $B$ } 
\begin{figure}[h]
\begin{center}
\includegraphics[width=0.48\textwidth]{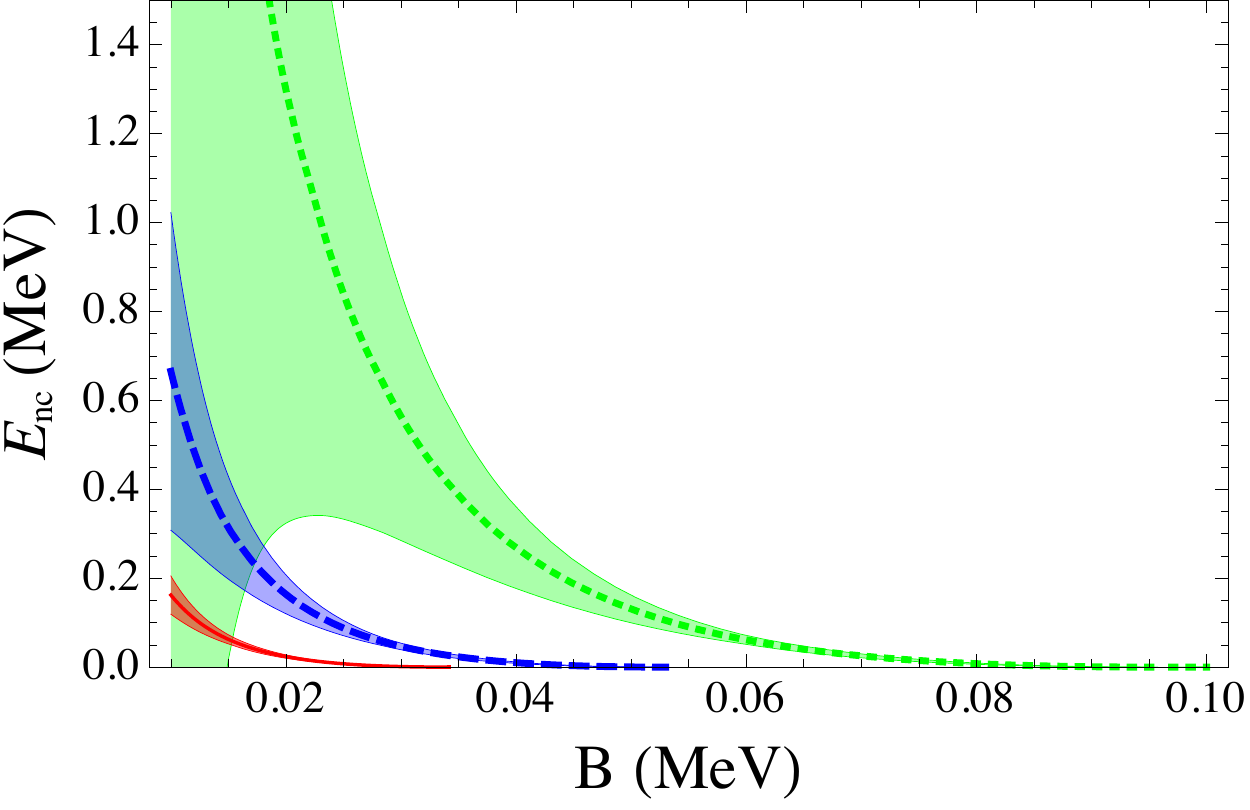}
\caption{Plots of $\sqrt{\langle r^2 \rangle} $~=~5.4~fm (blue, dashed), 6.3~fm (red, solid), and 4.5~fm (green, dotted), with their theoretical error bands, in the $(B,E_{nc})$ plane.}
\label{fig:contourplots}
\end{center}
 \end{figure}

In Figure~\ref{fig:contourplots}, we plot the sets of  ($B$, $E_{nc}$) values that give $\sqrt{<r^2>}=$ 4.5~fm, 5.4~fm and 6.3~fm, along with the theoretical error bands.  All sets of $B$ and $E_{nc}$ values in the plotted region that lie within the area bounded by the edges of these bands  give an rms matter radius within the combined (1-$\sigma$) experimental and theoretical error of the value extracted by Tanaka et al.. The figure shows that, regardless of the value of the $^{20}\text{C}-n$ virtual energy, Tanaka et al.'s experimental result puts a model-independent upper limit of 100~keV on the 2$n$ separation energy of $^{22}\text{C}$. 
 
Since Yamashita et al. obtain an LO matter radius that is too large for a given binding energy, their constraints on  the maximum possible value of $B$ are about 20\% weaker than ours. We also reiterate that their results are based on an incorrect assumption regarding the high-energy behavior of the integral-equation kernel. Our result for the maximum binding energy of ${}^{22}$C is a factor of two smaller than that found by Fortune and Sherr~\cite{Fortune:2012zzb}.  Ref.~\cite{Fortune:2012zzb} postulated that a simple extension of the model-independent relationship between $B$ and $\langle r_0^2 \rangle$ that prevails in a one-neutron halo~\cite{Hammer:2011ye} applies to two-neutron halos. Our calculation suggests that the relationship proposed in Ref.~\cite{Fortune:2012zzb} does not accurately capture the three-body dynamics of the 2$n$-halo system.
 
 \subsection{Implications for existence of excited Efimov states } 
\begin{figure}[h]
\begin{center}
\includegraphics[width=0.48\textwidth]{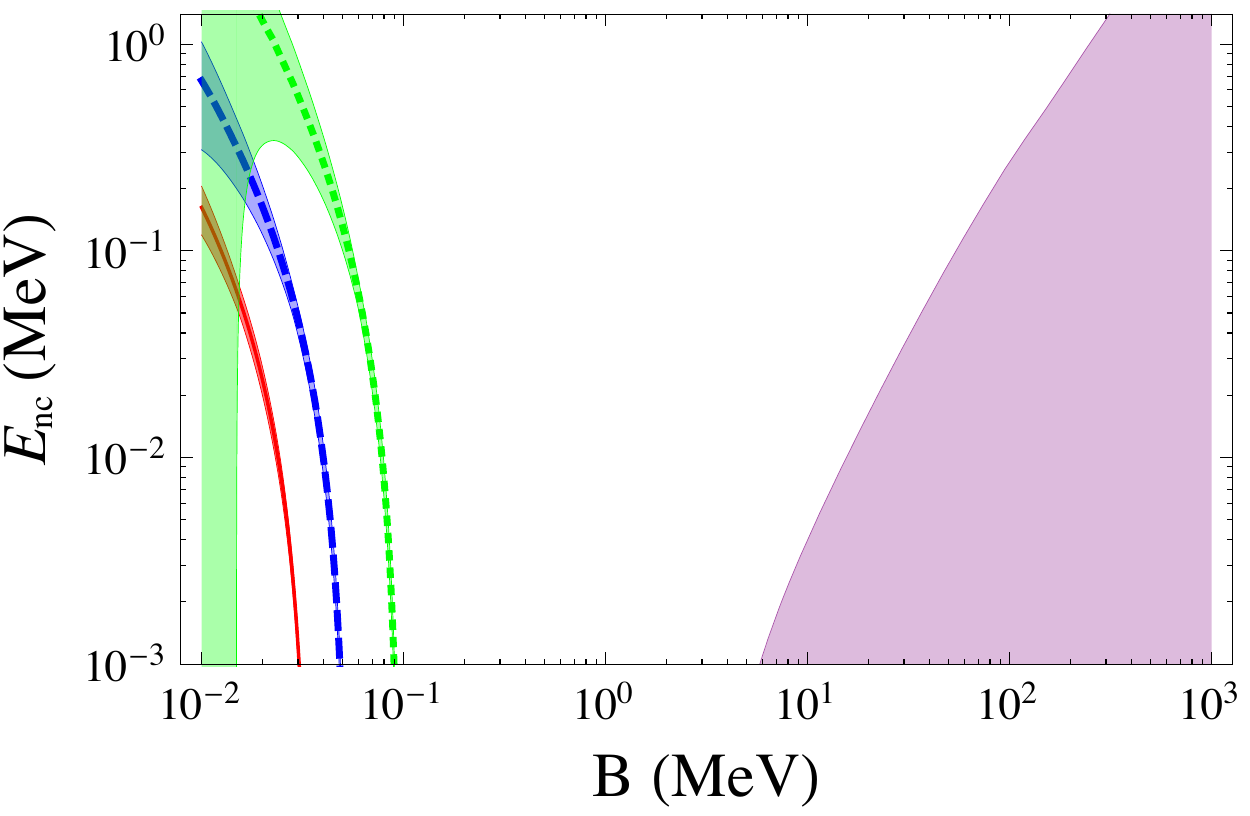}
\caption{The region in the ($B,E_{nc}$) plane that allows excited Efimov states (purple), and the region that encloses values  consistent with the experimental rms matter radius of $5.4\pm 0.9$~fm (with same color-coding as in Figure~\ref{fig:contourplots}).}
\label{fig:encvsblog}
\end{center}
 \end{figure}

Following Refs.~\cite{Canham:2008jd,Canham:2009xg}, we construct a region in the ($B,E_{nc}$) plane within which an excited Efimov state in ${}^{22}$C could occur. In Fig.~\ref{fig:encvsblog} the purple region is that which allows at least one excited Efimov state above the ground state in an $A$=20 Borromean nucleus. In the same plot, we also show the boundary curves that enclose the sets of  $E_{nc}$ and $B$ values which are consistent with an rms matter radius of $5.4\pm0.9$~fm once the theoretical errors are taken into account, i.e. those already displayed in Fig.~\ref{fig:contourplots}. 
 
The Efimov-excited-state-allowed and rms-radius-constraint regions do not overlap for a $^{20}\text{C}-n$ virtual energy larger than a keV. (The ${}^{22}$C radius can be computed accurately for ${}^{20}$C$-n$ virtual-state energies very close to threshold, but the computation of the existence of an Efimov state becomes numerically delicate here.) 
Indeed, as long as the trend in Fig.~\ref{fig:encvsblog} continues, Efimov states seem to be precluded for values of $E_{nc}$ well below 1 keV. However, we cannot make a stronger statement than this, since an Efimov state {\it is} present if we take $E_{nc}=0$, as a consequence of the fact that there are two ${}^{20}$C$-n$ pairs in the ${}^{22}$C system~\cite{Braaten:2002}. 
Therefore, while we cannot categorically rule out the existence of an Efimov state in ${}^{22}$C, we can say that the ${}^{21}$C system would need to be tuned very close to the unitary limit in order for one to be present.

 \section{Conclusion} 
\label{sec:conc}
We used Halo EFT at leading order to examine the behavior of the rms matter radius of s-wave Borromean halo nuclei in the $(E_{nn}/B,E_{nc}/B)$ plane. We computed the universal function $f$ which describes this behavior in a model-independent fashion. We then applied these results to ${}^{22}$C, and put constraints on the $(B,E_{nc})$ parameter space using the experimental value of the ${}^{22}$C matter radius. In contrast to previous works~\cite{Fortune:2012zzb,Ershov:2012fy,Yamashita:2011cb} which examined this problem, our constraint makes very few assumptions about the structure of ${}^{22}$C. We use only the experimentally well-supported idea that it can be treated as a three-body system composed of ${}^{20}$C and two neutrons. Furthermore, our result incorporates the anticipated theoretical uncertainty of the leading-order Halo EFT calculation based on this cluster picture. Even after this uncertainty, and the experimental (1-$\sigma$) error bar, are taken into account we find that $B<100$~keV for all values of $E_{nc}$. This rules out the possibility of an excited Efimov state in the $^{22}\text{C}$ nucleus unless the $^{20}\text{C}-n$ system has a virtual state with an energy much smaller than 1 keV.

This work was supported by the US Department of Energy under grant DE-FG02-93ER40756 and also in part by both the Natural Sciences and Engineering Research Council (NSERC) and the National Research Council of Canada. We thank Charlotte Elster and Hans-Werner Hammer for useful discussions. We are also grateful to Tobias Frederico, Hans-Werner Hammer, and Lucas Platter for useful comments on the manuscript.

 \end{document}